\documentclass[a4paper]{article}

\usepackage[english]{babel}
\usepackage{RR}
\RRNo{6183}
\usepackage{graphicx}
\usepackage{times}
\usepackage{caption}
\usepackage{verbatim}

\usepackage{graphicx}

\RRdate{2007}

 \RRdater{Avril 2007}
\RRNo{6183}

\RRauthor{
	Nathalie Henry \thanks{INRIA/LRI Univ. Paris-Sud \& Univ. Sydney}
  \and Jean-Daniel Fekete
  \and Michael J. McGuffin \thanks{Ontario Cancer Institute}
  
}
\authorhead{Henry \textit{et al.}}
\RRtitle{NodeTrix: Représentation Hybride pour Analyser les Réseaux Sociaux}
\RRetitle{NodeTrix: Hybrid Representation for Analyzing Social Networks}

\title{NodeTrix: Hybrid Representation for Analyzing Social Networks}

\titlehead{NodeTrix}
\RRresume{Alors que les données issues des communications electroniques
deviennent de plus en plus facilement accessible et comportent des informations
toujours plus riches et nombreuses, le besoin de visualiser ces réseaux sociaux
se fait plus pressant.  Cependant, les systèmes actuels ne permettent pas aux
analystes de}

\RRabstract{The need to visualize large social networks is growing as
hardware capabilities make analyzing large networks feasible and many new data
sets become available. Unfortunately, the visualizations in existing systems do
not satisfactorily answer the basic dilemma of being readable both for the
global structure of the network and also for detailed analysis of local
communities.  To address this problem, we present NodeTrix, a hybrid
representation for networks that combines the advantages of two traditional
  representations: node-link diagrams are used to show the global
  structure of a network, while arbitrary portions of the network can
  be shown as adjacency matrices to better support the analysis of
  communities.  A key contribution is a set of interaction techniques. These
 allow analysts to create a NodeTrix visualization by dragging selections from
 either a node-link or a matrix, flexibly manipulate the NodeTrix representation
 to explore the dataset, and create meaningful summary visualizations of their
 findings.   Finally, we present a case study applying NodeTrix to the analysis
 of the InfoVis 2004 coauthorship dataset to illustrate the capabilities of
 NodeTrix as both an exploration tool and an effective means of communicating
 results.  }

\RRmotcle{Visualisation de graph, Réseaux sociaux, visualisation matricielle,
visualisation hybride, aggregation, interaction.}

\RRkeyword{Graph visualization, Social networks, Matrix visualization, Hybrid
  visualization, Aggregation, Interaction.}

\RRprojet{AVIZ}

\RRtheme{\THCog} 

\URFuturs 

\begin{document}
\makeRR   

\section{Introduction}

Social network analysis is a growing area of the social sciences.
Vast new datasets are becoming available as people conduct ever more
of their social lives electronically.  Online projects such as
Wikipedia or open-source software development are creating new social
networks on a global scale.  At the same time, the challenges of a
more integrated world generate new demands for analysis such as
monitoring terrorist networks or the spread of potentially pandemic
diseases.  Social network visualization is becoming a popular topic in
information visualization, generating more and more tools for the
analysts.  In 2006, 10 network-related articles have been presented at
the InfoVis Symposium (30\% or the articles) and 6 at the VAST
symposium.  The large majority of the network visualization systems
use the node-link representation: 54 (out of 55) node-link based
systems referenced in the Social Network Analysis
Repository (http://www.insna.org/)%
, and 49 (out of 52) on the Visual
Complexity website%
(http://www.visualcomplexity.com/).  This
representation is well suited to show sparse networks, but social
networks are known to be globally sparse and locally dense.
Therefore, social network visualization faces a major challenge:
obtaining a readable representation for both the overall sparse
structure of a social network and its dense communities.

\begin{figure}[ht]
\centering
\includegraphics[width=8.5cm]{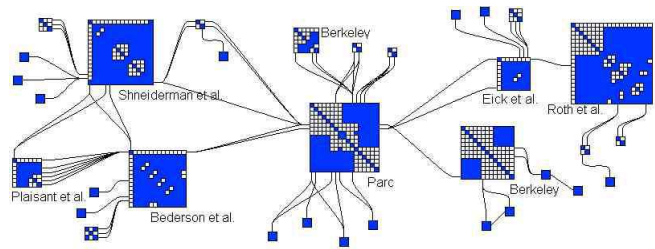}
\caption{NodeTrix Representation of the largest component of the InfoVis
Co-authorship Network}
\label{fig:infoviscompact}
\end{figure}

In this article, we propose a novel visualization called
\emph{NodeTrix} to address this challenge.  NodeTrix integrates the
best of the two traditional graph representations by using node-link
diagrams to visualize the overall structure of the network, within
which adjacency matrices show communities. 

 
The article is organized as follows: after the related work section,
it describes the NodeTrix representation and the data structure it
relies on.  It then details the interaction techniques we designed for
creating a NodeTrix hybrid, either by starting from a standard
node-link diagram or from a standard adjacency matrix.  Finally, it
describes a case study using NodeTrix to explore and present the
results of a co-authorship social network.

\section{Related Work}

\subsection{Social Network Analysis}

Social networks are graphs, where the nodes are actors (people) and
the edges are relationships.  They vary from very sparse (genealogical
trees) to very dense (exports and imports between countries).
\emph{Small-world networks} belong to an intermediate category that
occur very frequently in social networks, including many
acquaintanceship networks as well as the global Internet.  They are
the focus of many studies~\cite{ComplexNetworks,cscw2004} because of
their interesting properties~\cite{SmallWorld}.  For social network
visualization, the most relevant of these properties are a high
clustering coefficient, corresponding to the presence of many
\emph{locally dense} clusters, and a small cross-section, caused by a
small number of hub nodes connecting a graph that is \emph{globally
  sparse}.

Social network analysis relies on three important tasks~\cite{SocNetAnal,Scott}:
\begin{itemize}
\item identify \emph{communities}, \textit{i.e.} cohesive groups of
  actors that are strongly connected to each other;
\item identify \emph{central actors}, \textit{i.e.} actors linked to
  many others or that bridge communities together;
\item analyze roles and positions --- these are higher level tasks relying on the
  interpretation of groups of actors (positions) and connection patterns (roles).
\end{itemize}

We now consider each of these three tasks in more detail, pointing out
the corresponding graph-theoretic properties or graph analysis tasks
using the taxonomy of tasks in~\cite{GraphTaskTaxonomy}.

To perform \emph{community analysis}, an analyst should be able to
group actors by attributes and to study the connection patterns within
each group.  The analysis of attributes such as actors' names or
interests is important to label the community in question and
interpret why these actors are grouped.  Studying the connection
patterns reveals how actors of the community are linked and the
strength of their relationships.  Analysts need to evaluate the
density of a community in terms of connections and also to quickly
identify \emph{cliques} (a group where each actor is linked to every
other actor) and missing relationships.  Thus, community analysis
relies on attribute-based tasks, involving attributes of actors or
relationships, and topology-based tasks,
such as examining adjacency (direct connections) between nodes.

\emph{Identifying central actors} is revealed by performing
essentially topology-based tasks.  Analysts need to identify the most
connected actors, as well as articulation points (actors bridging
communities together).  Such actors can be identified using measures
of centrality, several of which are based on path-related tasks.  For
example, the betweeness centrality indicates the number of times a
node is present in a shortest path between every pair of nodes in the
network.  Identifying central actors requires understanding the global
structure of the network, \textit{i.e.} finding communities, how they are
linked and what actors link them.

\emph{Analyzing roles and positions} is done by analyzing how actors
are connected within a community and outside a community.  This task
requires more interpretation and relies also on attributes of actors
and relationships.

Many systems exist to analyze social networks.  We classify them into two
categories: menu-based systems and exploration systems.

Menu-based systems provide a wide range of functionalities but users
often needs expert help or a cookbook to analyze their datasets.
Examples of these systems include Ucinet~\cite{UCINET} --- based on
statistics and proposing a broad range of analysis functions --- and
Pajek~\cite{Pajek} which provides a large set of algorithms to
partition, permute, cluster, hierarchize and layout networks.

Mastering all the functionality of these menu-based systems to control
the analysis process requires considerable effort from the user, hence
recent systems are aimed at a more exploratory process.  This process
is based on starting with an overview of the whole network and then,
using interactions (MatrixExplorer~\cite{Henry:2006:TMV}) or simple
scripts (Guess~\cite{Guess}), to manipulate the dataset (\textit{e.g.}
through filtering or clustering) and create a set of visuals for
further analysis.  A number of recent systems forgo the first step of
this process because displaying a readable overview of a large graph
is too difficult.  PivotGraph~\cite{PivotGraph} proposes to start the
exploration from a top level aggregation of network attributes.  The
user visualizes categories and their relationships, and then interacts
with the visualization to explore lower levels.
NetLens~\cite{NetLens} focuses on simple visualizations (histograms)
of the network attributes and interaction to analyze the network.
Semantic substrates~\cite{SemanticSubstrates} relies essentially on
filtering and organizing actors according to their attributes.
Finally, TreePlus~\cite{TreePlus} and Vizster~\cite{Vizster} focus on
a local representation of the network and use interaction to navigate
within the whole dataset.  These systems present interesting ideas and
are very comfortable to use.  However, identifying central actors and
understanding the global structure of the network using these systems
is difficult.

\subsection{Graph Drawing}

Graph drawing has a very rich history~\cite{dibattista1999,survey},
with early work on interactive computerized visualizations extending
back to the 1960s~\cite{baecker1967}.  However, almost all
visualizations of graphs amount to either node-link diagrams or
adjacency matrix representations.  There are a few examples of hybrid
representations for graphs~\cite{harel1988,sindre1993} and for trees
\cite{zhao2005,fekete-curvedlinks2003} but they do not combine
node-line diagrams with adjacency matrices.

Node-link diagram is the most familiar representation of graphs in
general and social networks in particular.  It is good at showing the
overall structure of a sparse graph, but Ghoniem et
al.~\cite{GraphReadability} showed that density has a strong impact on
its readability.  Focusing on basic readability tasks such as finding
an actor or determining if two actors are linked, they conclude that
node-link diagrams perform badly for dense graphs even with few
(\textit{e.g.} 20) nodes.  Because node-link diagrams become
unreadable in dense communities and around high-degree hub nodes, they
do not lend themselves to community analysis.

(As an aside, in a community that is almost a clique and only missing
a few edges, it might be suggested to use a ``complementary''
node-link diagram, where the links displayed indicate the {\em
  missing} edges; all the other edges being implicitly present.  This
would reduce clutter in some case, but in general is not a viable
solution, because a community of $n$ nodes within $n^2/4 = O(n^2)$
edges is considered dense, but has an approximately equal number of
direct and missing edges.  Thus, clutter remains a significant problem
even with such ``complementary'' node-link diagrams.)

The adjacency matrix representation, which is particularly effective
for dense graphs, have been proposed to solve this
problem~\cite{MatrixZoom, Henry:2006:TMV}.  However, it is ill-suited
to path-related tasks~\cite{GraphReadability} that are very important
in social network analysis.  Analysts following paths between actors
using a node-link network representation can exploit ``Gestalt
continuation'' preattentive visual processing, but in a matrix
representation the same task requires aligning and matching nodes back
and forth between corresponding rows and columns, a tedious and
error-prone high-level cognitive task.  So, matrix representations
ease community analysis, but hinder identification of global
structure.

Thus, at one extreme are sparse social networks, which have an almost
tree structure and few communities, for which node-link diagrams are
well suited.  At the other end of the spectrum are very dense networks
in which matrices are well suited.  The problem boils down to deciding
which visualization is more suitable for small-world networks that
have an intermediate nature, being globally sparse but locally dense.
Choosing between these representations requires a trade-off between
readability of global structure and ease of community analysis.

Henry and Fekete~\cite{Henry:2006:TMV} chose to provide users with
both representations synchronized by selection.  They argue that users
can use the most appropriate visualization for each task.  However,
their system requires the use of two screens and the authors point out
the potential cognitive load and divided attention from switching
between representations.

Other recent work by Henry and Fekete~\cite{MatLink} attempts to
overcome the weaknesses of matrix representations by adding links on
the sides of the traditional matrix.  While the authors experimentally
demonstrated that their visualization improves the traditional matrix,
their results also show that the user fails to identify some important
features (in particular, the articulation points) of networks.

Solutions have also been proposed to improve the readability of
node-link diagrams for communities.  Auber et al.~\cite{SmallWorldVis}
introduce aggregated node-link diagrams where each community is
aggregated in a single node within which a small overview is
displayed.  While communities are quickly identifiable and the global
structure more readable, detailed analysis of communities is
impossible because links between communities are missing.

Holten proposed the Hierarchical Edge Bundles
technique~\cite{EdgeBundle} to improve the readability of hierarchical
graphs; it can also be applied to clustered graphs.  Although it can
improve the readability of the global structure and inter-community
relationships, it is still difficult to identify intra-community
organization as nodes inside clusters are positioned along a circle,
creating many edge crossings.

\section{NodeTrix}

NodeTrix is a hybrid representation of networks based on the node-link
diagram where communities can be represented as matrices.
Intra-community relationships use the adjacency matrix representation
while inter-community relationships use normal links.

\subsection{Data Structure and Design Choices}
Two graphs are involved in a NodeTrix representation: the raw {\em
  underlying} graph (composed of underlying nodes and edges) that
serves as initial input, and an {\em aggregated} graph (composed of
aggregated nodes and edges) that is derived from the underlying graph.  
Each aggregate node may correspond
to either a unique underlying node or to a group of underlying nodes
that typically form a community.  Underlying nodes are never shared by
aggregate nodes, \textit{i.e.} there is a many-to-one mapping from
underlying nodes to aggregate nodes (and also from underlying edges to
aggregate edges.)

Because our goal with NodeTrix is to provide a readable representation
for dense subgraphs, only a single level of aggregation is used: dense
subgraphs are simply aggregated and displayed as matrices.  Some
aggregated nodes may correspond to only one underlying node rather
than a group of underlying nodes and these are displayed as a simple
node rather than a matrix.  However, operations are designed to be
uniform over all aggregated nodes.  In particular, the user can add or
merge aggregated nodes, whether each node involved corresponds to just
one or many underlying nodes.

Attributes of the underlying nodes and underlying edges are combined
and propagated up to the aggregated elements.  For nominal and
categorical attributes, values are combined through simple
concatenation.  Numerical attributes are aggregated either using the
average or the min,max values.  An interesting benefit of using
matrices in NodeTrix is that it can display the attributes of the
underlying elements and of the aggregated elements and that for both
the links and the nodes.  Furthermore, because users can dynamically
switch between the two representations, more visual variables are
available to show attributes.  For example, the background color of a
matrix can correspond to an aggregated node attribute, while
attributes of each underlying node can be shown along the axes of the
matrix.  Similarly, the axes can be used to display labels of
individual underlying nodes, while a global aggregated node label is
also shown.

\subsection{NodeTrix Visualization}

To render the NodeTrix representation, a standard node-link layout is
used for the aggregated graph, and in addition aggregated
nodes containing more than a single underlying node are overlayed with a 
matrix representation.  

\subsubsection{Drawing Matrices}

NodeTrix is built on the InfoVis Toolkit~\cite{InfoVis} and uses its
rendering mechanism to create the visualization.  The rendering
mechanism involves a pipeline of renderers which makes it simple to
draw a matrix over a standard node.  For example, a simple rendering
pipeline for a node-link diagram would be : compute\_position,
compute\_size, set\_color, fill\_shape, draw\_border, draw\_label.  To
overlay matrices on standard nodes, we introduced a matrix renderer
between the fill\_shape and the draw\_border renderers.  This renderer
displays the matrix after having rendered the background node (with a
given position, size and color) and before drawing the border used for
selection and the label.

Matrices have two advantages which make them more readable than
node-link diagrams to represent an aggregated node: first, as nodes
are organized linearly, edges from the rest of the graph to the
underlying nodes are readable and suffer from a limited number of
crossings; secondly, as nodes are represented both in rows and
columns, edges can be drawn from any of the four sides of the matrix,
which also limits crossings and overlapping problems.  Finally, rows
and columns of matrices can be reordered (manually or automatically)
to improve readability and further reduce the number of edge
crossings.

%

To save memory and allow the user to control all the matrices'
properties with a single general control panel, the matrix renderer
uses a single matrix visualization object, applying a different
permutation for each aggregated node.  Therefore, changing the color
attribute for the matrix axes will affect all displayed matrices.  We
considered creating a separate matrix object for each aggregated node
instead, allowing the user to display different attributes on
different matrices.  However, it would have been very confusing for
the user to manage all the controls in a single huge panel (one set of
controls for each matrix) or to force the user to select a matrix to
see its controls.  We decided that sharing the visual attributes for
all the matrices was the best compromise.

\subsubsection{Drawing Links}

To display links in NodeTrix, we considered three options: displaying
only aggregated links, displaying only the underlying links, or
displaying both.

\begin{figure*}[t]
\centering
\includegraphics[width=3cm]{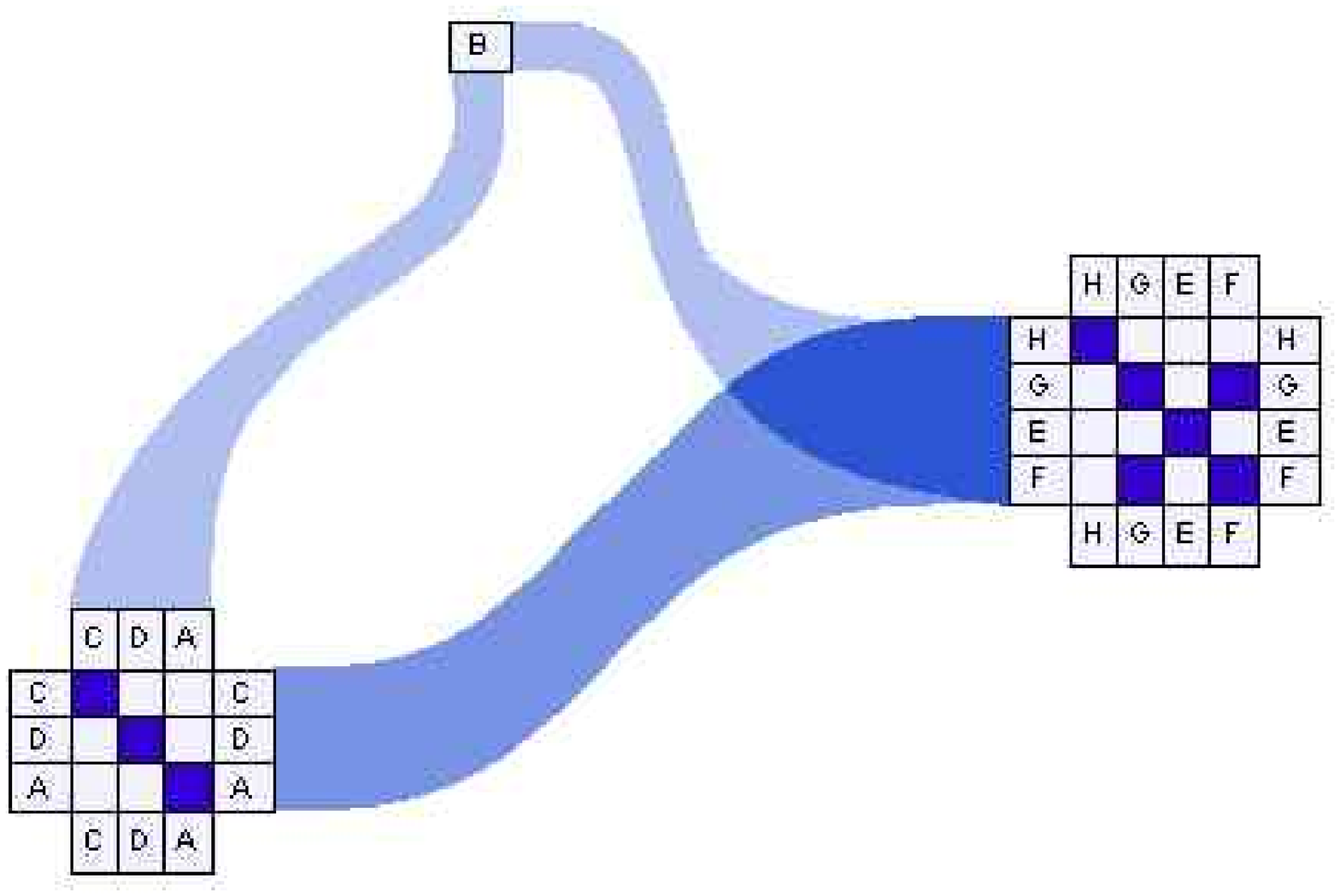}
\includegraphics[width=3cm]{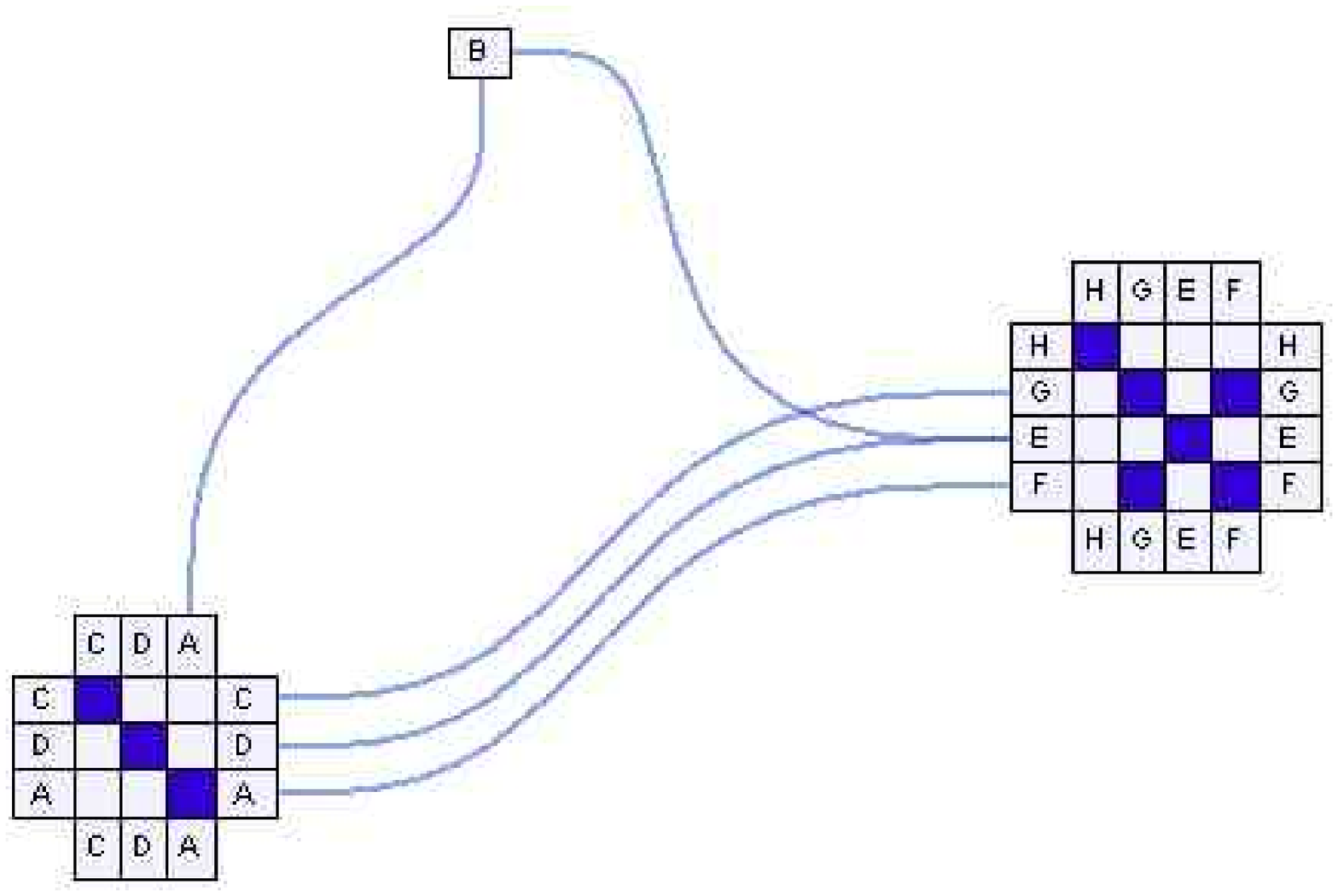}
\includegraphics[width=3cm]{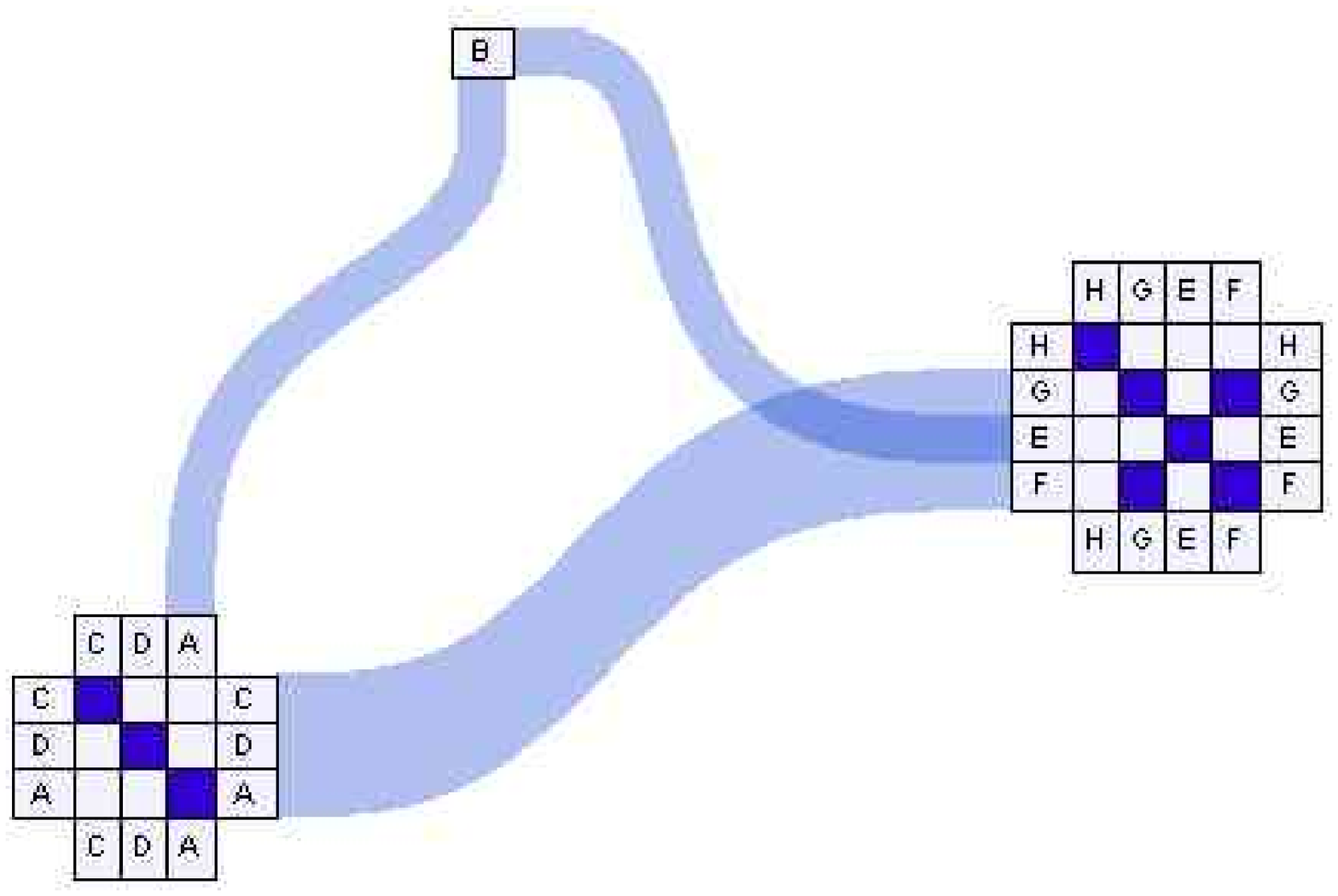}
\includegraphics[width=3cm]{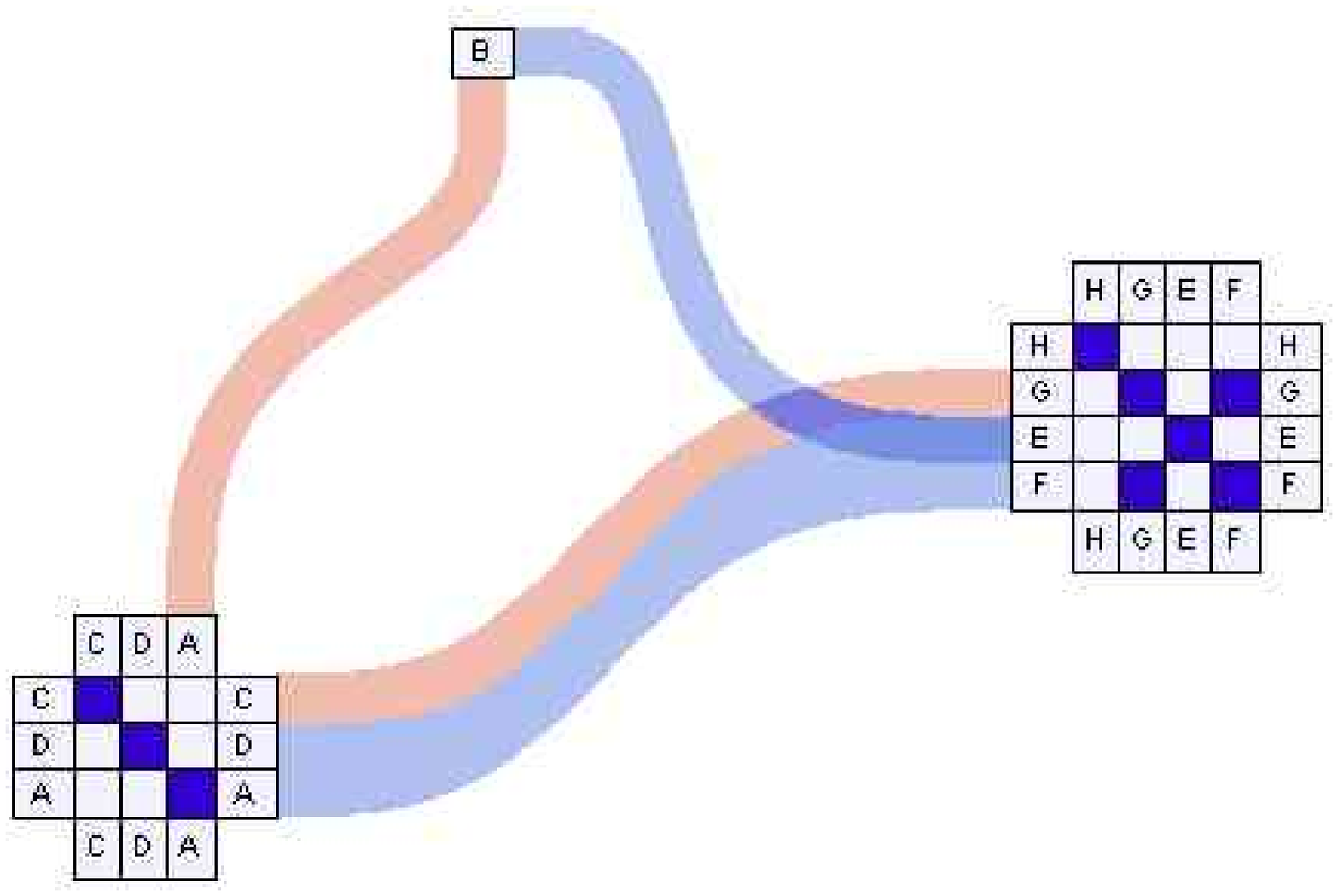}

\label{fig:edges}	
\caption{Drawing links (from left to right): (a)aggregated edges, (b)underlying
edges, (c)underlying edges with full size, (d)underlying edges carrying attributes}
\end{figure*}

Displaying aggregated links (Figure~\ref{fig:aggregatededges})
provides simple visual feedback on how communities interact.
Moreover, an aggregated attribute can be mapped to a visual variable
(\textit{e.g.} color, thickness, opacity) of this edge.  However, the
details of which actors of the two communities are interacting are not
visible.  On the other hand, displaying each underlying edge
(Figure~\ref{fig:edges}b) provides connectivity details and enables
visualization of the attributes of each edge independently, but at the
cost of many more links and potential crossings.  Because small-world
networks are globally sparse, they are few inter-community
relationships.  However, displaying both aggregated and underlying
edges at a same time could nevertheless be confusing, in part due to
the possible interaction between visual variables and edge crossings
or overlap.

For NodeTrix, we chose to visualize underlying edges, but with the
added flexibility of allowing the user to control the thickness of
links through a slider.  Increasing the size of underlying links
eventually causes them to merge, and the resulting visual feedback
(Figure~\ref{fig:edges}c) is similar to that with aggregated
edges (Figure~\ref{fig:edges}a), but with more precision.  Moreover,
when visualizing an underlying edge attribute as a color, the
thickness of merged bands of color conveys the number of underlying
edges (\ref{fig:edges}d).  The slider that controls
links size updates the visualization with smooth, immediate feedback,
similar in spirit to direct manipulation.  Manipulating this slider
allows the user to quickly switch from one kind of overview mode ---
how are communities linked?  Which kinds of links? --- to a detailed
mode --- who are linking the communities together?

\subsubsection{Layout}

Because the aggregated graph in NodeTrix is laid out as a traditional
node-link diagram, any existing graph layout could be used.  However,
because NodeTrix is intended to be used as an interactive exploration
tool and we do not want to confuse the user with large, sudden changes
to the layout, it seems appropriate to support incremental,
interactively-driven changes to the layout, such as aggregating or
splitting nodes.  The initial layout given to the graph is the LinLog
layout proposed by Noack~\cite{LinLog}.  It was chosen to give
prominence to clusters so they can be quickly identified by the user.
After this initial layout, the user may then make local changes such
as dragging nodes to change their positions, grouping a set of nodes,
or removing a node from a group.

To (re)order the nodes within an adjacency matrix, many different
algorithms can be used.  As these matrices are typically small, the
running time is not an issue.  Nevertheless, we chose not to reorder
the matrices automatically as they are usually very dense and do not
need any particular optimization.  Instead, we preferred to allow the
user to interactively move rows and columns.

\subsection{Visual Variables and Control Panel}

NodeTrix relies on the InfoVis Toolkit to generate controls to filter
and affect visual variables.  The user controls two sets of visual
variables: one for the node-link diagram, and one for the matrices
displayed in the aggregated nodes.  Each set of variables contains the
following, for nodes and edges: color, transparency, shape size,
filled area of the shape, border color, width, and labels.

The user filters and associates visual variables to aggregated and
underlying graph attributes using simple controls such as combo boxes
or sliders.  The visualization is immediately updated, following the
principles of direct manipulation.

\section{Interaction}

We designed a set of interaction techniques to create, edit and
manipulate NodeTrix in a very simple and powerful way because we
believe that manipulation is key to understanding a network and its
potential multiple interpretations.

\subsection{NodeTrix Editing}

NodeTrix can be created starting with a pure, traditional node-link
diagram.  We propose a set of interactions based on drag-and-drop of
nodes, matrix axis items, and matrix core elements.  We feel these
interactions are easy to understand as the user simply grabs one of
these elements and drops it to another location (possibly over
existing elements) to perform an action.  When dragging an element,
the user has immediate visual feedback and is able to read the
element's label.

\emph{Moving a node or a matrix} to adjust its position and improve the
readability of the representation can be done by grabbing the matrix or the
node, dragging it and releasing it at a new position.  As the element is
dragged, its connecting links are updated.

To \emph{aggregate a group of nodes} into a matrix, the user may
lasso-select the desired nodes, which are then immediately converted
into a matrix.  To make the transition to a matrix smooth, the
transformation from node-link diagram to matrix is animated.  The
animation speed is adjustable to suit both novice users (who may
benefit from seeing a slow animation, to better understand how nodes
and edges become organized into a matrix) and advanced users (who
would presumably prefer a brief animation).
\emph{Splitting a matrix} back into a group of nodes is done by
right-clicking on it, in which case nodes are positioned with a circular layout
around the center of the previous matrix.

To complete these basic aggregation features, we provide additional
interactions for finer-grain editing of the aggregated elements.  If
users missed an element with the lasso selection or simply wants to
\emph{add an additional node} to a matrix, they can drag-and-drop a
single node into the matrix.  The node will integrate with the matrix,
appearing in the matrix axis items (in both the rows and columns).
Its connections with the matrix elements will be displayed in the
matrix core, whereas its connections with the external elements will
be displayed as links starting from the matrix axis items and ending
at the external elements.  If a single node is dragged onto another
single node, then the two will be aggregated into a $2\times 2$ matrix.
On the other hand, if users wish to \emph{extract a node} from a matrix, they
can grab the corresponding matrix axis item (either on the row or column
axis) and drop it outside the matrix.  The dropped item will then be displayed as
a standard node with appropriate links between itself and the matrix, and the 
corresponding row and column in the matrix will be removed.  

To increase readability or visualize different combinations, users
may want to \emph{move an item} from one matrix to another.  This can
done by grabbing a matrix axis item and dropping it on the other
matrix.  During the transfer, the user is able to read the node label
and may cancel the interaction by dropping the element back into the
original matrix.  This may result in a change to the ordering of nodes
in the matrix.  The \emph{order of items} in the matrix normally
corresponds to the item addition order, with the last item added in
the last position.  However, when two matrices are merged, the item
ordering follows the indices of nodes in the underlying graph.  The
ordering of nodes can be changed by grabbing nodes and dropping them
back into the matrix, one at a time, in the desired order.

Finally, users can \emph{merge matrices} together by
dragging-and-dropping a matrix over another.  

\subsection{Geometric Zoom on Matrices and Axis Labels}

An aggregated matrix may occupy more space than the original group of
nodes in node-link representation.  This is partly due to the labels
displayed on each side of the resulting matrix.  However, while
reading labels on each side of the matrix is required to perform
community analysis and local editing operations, the axis labels are
not required on all matrices at all times, and the size of the matrix
core can be reduced to fit the minimum level of readability.
Moreover, as each matrix possesses a label (reflecting its
composition), axis labels for individual underlying nodes may not be
necessary at all in a final layout.

We tried displaying the axis labels on demand following the excentric label
principles.  For example, if the mouse pointers hovers over a matrix, its
axis labels as well as its neighbors' axis labels are 
displayed.  In this case, axis labels
need to remain visible after the mouse pointer moves (to avoid frustrating the user
when loosing a landmark or pointing at an item to grab).
However, during a case study, we observed that it was more
comfortable to be able to read all axis labels when editing, and to
remove all axis labels at once and reduce the size of the matrices to
get an overview of a final layout.

For these reasons, we added two sliders in the control panel to
control the size of the matrices and the axis labels.

\subsection{Supporting the Exploration of Matrices}

One strong weakness of the matrix representation, when exploring a
network, is the tedious work required to perform path-related tasks.
For example, finding how two communities are connected is tedious as
it requires going back and forth alternately reading rows and columns.
Moreover, if communities are far apart in the matrix, this task
requires a scan of the full length of matrix rows or columns, and
connections in a large matrix may lie outside the viewport.
Obviously, the task is worse when dealing with three matrices as the
user needs to check for intersections of rows and columns in each of
the three communities.

We noticed in a participatory-design session reported
in~\cite{Henry:2006:TMV} that social network analysts also use the
matrix representation for some of their analyzes.
To help perform community analysis and provide support for
path-related tasks in general, we provide users with a couple of
interaction techniques that work across separate matrix-NodeTrix
windows, that might be arranged in a dual-viewport or split-screen
fashion.  These techniques are still based on drag-and-drop, however
this time, the user drags a group of elements from one window to
another one.

The interaction is made of two steps: first, the user selects a group
of nodes in the window of the pure matrix visualization and then drags
this group to the NodeTrix window.  To select the group of nodes, we
provide lasso selection directly on the pure matrix representation.
Alternatively, the selection can be done on an axis (rows and
columns).  When a group of edges (matrix cells) is selected, the
corresponding set of nodes transferred is the union of the edges'
source nodes and sink nodes.  
Dropping the selected group inside the NodeTrix window performs the
addition of an aggregated node to the NodeTrix visualization.  The
group is then displayed as a matrix.  Selecting and dropping a second
group allows the user to see how these groups are connected to each
other visualizing the result with links.  The process can continue
to visualize connections between several communities.


\section{Animation}

Proper use of animation has much potential to increase the
effectiveness of user interfaces and visualizations
\cite{woods1984,baecker1990,bartram1997}.
To help users maintain their mental model of the
network across interactions, we considered how to
continuously animate the aggregation of nodes into
an adjacency matrix.
Typically, animating over transitions involves some kind of
interpolation of graphical elements from one state to another.
In the case of transitioning from a node-link diagram to
a matrix, however,
the visual design of the animation is non-trivial,
because node-link diagrams and adjacency matrices are
composed of very different graphical elements.
There is a sort of duality between the two forms:
nodes correspond to {\em points} in node-link diagrams,
but to {\em line segments} (rows and columns) in matrices,
and, conversely, edges correspond to {\em line segments} in node-link diagrams,
but to {\em points} (intersections of rows and columns) in matrices.
The key problem is to find an intermediate graphical form or layout
through which we can interpolate during an animation.

\begin{figure}[ht]
\centering
\includegraphics[width=8cm]{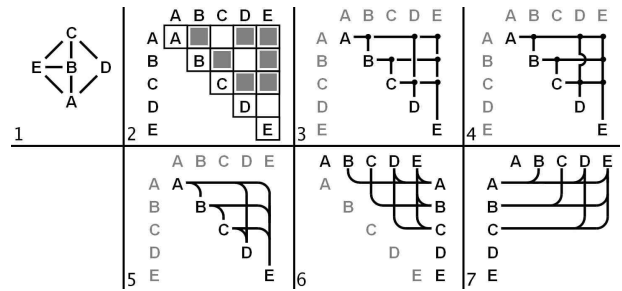}
\caption{1: A node-link diagram of a network. 
2: The corresponding adjacency matrix.  For simplicity, only the upper half
is shown, since the matrix is symmetric.
3 through 5: different ways of depicting the edges in a node-link diagram
laid out over the matrix, using polylines or curves.  The ``corners'' of the edges coincide with the filled-in cells of the matrix in 2.
3 and 4: inspired by circuit wiring diagrams.
5 through 7: different choices for the locations of nodes in the node-link
diagram laid out over the matrix.
6 and 7: each node is duplicated and has two locations in the node-link 
diagram. }
\label{fig:animation1}
\end{figure}

To find solutions,
we conducted sessions of sketching, brainstorming, and analysis of how
graphs can be depicted with node-link diagrams and matrices.
We noticed that, although each node corresponds strictly to an entire row and
column within a matrix, the node can also be identified with special
points in the matrix, that occur where the diagonal and the axes (or sides)
of the matrix intersect the node's row and/or column.
Furthermore, it is possible to draw a node-link diagram overlaid on a matrix grid,
in such a way that the nodes fall on some of these special points,
and such that the edges
(drawn as polylines or curves) pass through their own corresponding
locations in the matrix.
Figure~\ref{fig:animation1}, subfigures 3--7, show some possibilities.

As can be seen, there are several possibilities for the intermediate state
that an animation might interpolate through.
We identify a few different design dimensions.
First, the edges in the intermediate state might be depicted
using polylines or curves (Figure~\ref{fig:animation1}, subfigures 3--5).
Second, the location of nodes might be along the diagonal
or along the sides of the matrix (subfigures 5--7);
in the latter case, each node must be duplicated at some point during the animation.
Third, the intermediate state might show only the upper half of the matrix
(after which the animation might fade in or unfold the other half of the matrix
as a mirror image), or the intermediate state might show the whole matrix
(before which the animation would have to duplicate the edges somehow, since they
occur in each half of the matrix).

\begin{figure}[ht]
\centering
\includegraphics[width=8cm]{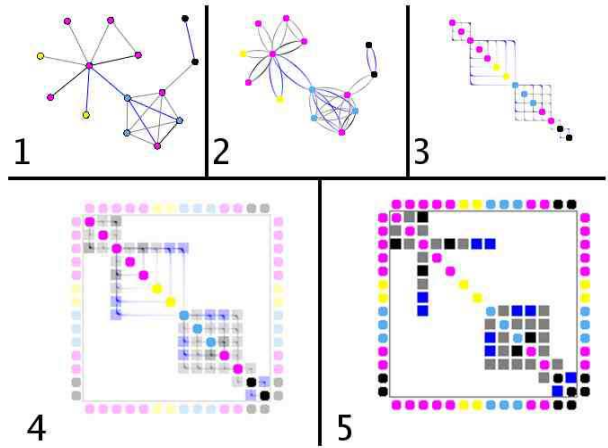}
\caption{The stages of an animation from a node-link diagram (1) to an adjacency matrix (5).
}
\label{fig:animation2}
\end{figure}

We made a first set of choices along each of these design dimensions
and implemented an animated transition from node-link diagrams to
adjacency matrices, both in the NodeTrix software and in an additional
piece of software.  Figure~\ref{fig:animation2} shows the latter
implementation, where the network has colored nodes and edges.
As can be seen, the intermediate state (subfigure 3) shows both
halves of the matrix, hence the animation begins by
duplicating edges (subfigure 2).  The positions of the nodes,
and of the control points for the edge curves,
are gradually interpolated to reach their final locations (subfigure 3).
Then, the edge curves are faded out as the normal depiction
of the matrix is faded in (subfigure 4).
Notice that the ``corners'' of the edge curves coincide with
the appropriate cells of the matrix (subfigure 4),
and the opacity of the curves is varied such that these corners
are the last part of the curve to fade away, to reinforce
their visual correspondence to the matrix cells that fade in.

Compared with other animated transitions in visualization systems,
this animation may seem rather complicated, and in practice
an expert user may prefer that the animation be brief
(e.g.\ lasting 0.5 seconds).  However, novice users may appreciate
having these animations last longer, at least initially.
We expect that, in addition to helping the user maintain
a mental model of the visualization across transitions,
these animations may also have an educational benefit,
to help users learn how adjacency matrices are constructed
and how to interpret them.
We expect it would be worthwhile to implement variations on the animation
corresponding to the other design choices we identified,
and to solicit feedback from users as to their preferences.

A fourth design dimension relevant for
education involves deciding whether to animate all the nodes
and edges at the same time, or to animate them in sequence.
For example, edges might be animated one at a time, constructing
the matrix cell-by-cell, or alternatively, each node (with all its edges)
might be animated one at a time, constructing the matrix
row-by-row.  Such a sequential animation might be made to accelerate
as more of the matrix is built-up, allowing the user to see
the process in detail at first, and then to see it quickly
complete the rest of the matrix.

\section{Case Study: Exploring and Presenting Publications Data}

In this case study, we present how NodeTrix can be used both for
exploring and presenting publications data.  The InfoVis 2004 contest
provided us with a clean dataset from which we extracted the
co-authorship network of the Information Visualization field.  This
network is disconnected in 291 components and contains 1104 nodes
(researchers) and 1787 edges (co-authorship).  It has a low density
and a high clustering coefficient, which categorizes this network as a
small-world network.

We only present here the analysis of the largest connected component,
containing 137 nodes and 328 edges.  This network can be considered as
small sized but it already presents challenges for exploration and
presentation using traditional matrix and node-link diagrams.
Detailed communities are not readable in node-links while connections
between communities are tedious to find in matrices.

Moreover, presenting results on paper requires some filtering for
matrices as well as node-link diagrams.  The matrix representation
requires space: it cannot fit in a printed article with readable
labels for networks of more than a hundred nodes.  Edge-crossings and
node-overlapping is an issue for node-link diagrams.  Filtering
reduces the size and density of the network to make it more readable.
Using NodeTrix solves these presentation problems since dense subgraph
are aggregated as matrices whereas edge-crossing and node-overlapping
is limited. Furthermore, communities (aggregated nodes) remain
readable.

\subsection{Setup}

To manipulate NodeTrix, we used an interactive pen display.  Pen-based
interactions on NodeTrix are intuitive and comfortable using this
input device.  The user can simply grab elements by pressing the pen
over them, drag them moving the pen on the screen and finally release
them by raising the pen.  Lasso selection provides also a very
intuitive feedback similar to the use of a real pen.

\subsection{Aggregation and Exploration}

NodeTrix is a flexible representation for which the level of
aggregation as well as the level of details is controllable.  For
example, Figure~\ref{fig:infovisnodetrix} and
Figure~\ref{fig:infoviscompact} show the exact same dataset: the
largest component of the InfoVis co-authorship network.  In the
compact representation (Figure~\ref{fig:infoviscompact}), the goal was
to provide a brief overview of main communities in the field, whereas
in the second representation, the goal was to be able to identify all
nodes of the network while grouping them by communities.  A third
representation could have been a more detailed representation with the
axes of the matrices displayed.

\begin{figure}[ht]
\centering
\includegraphics[width=3cm]{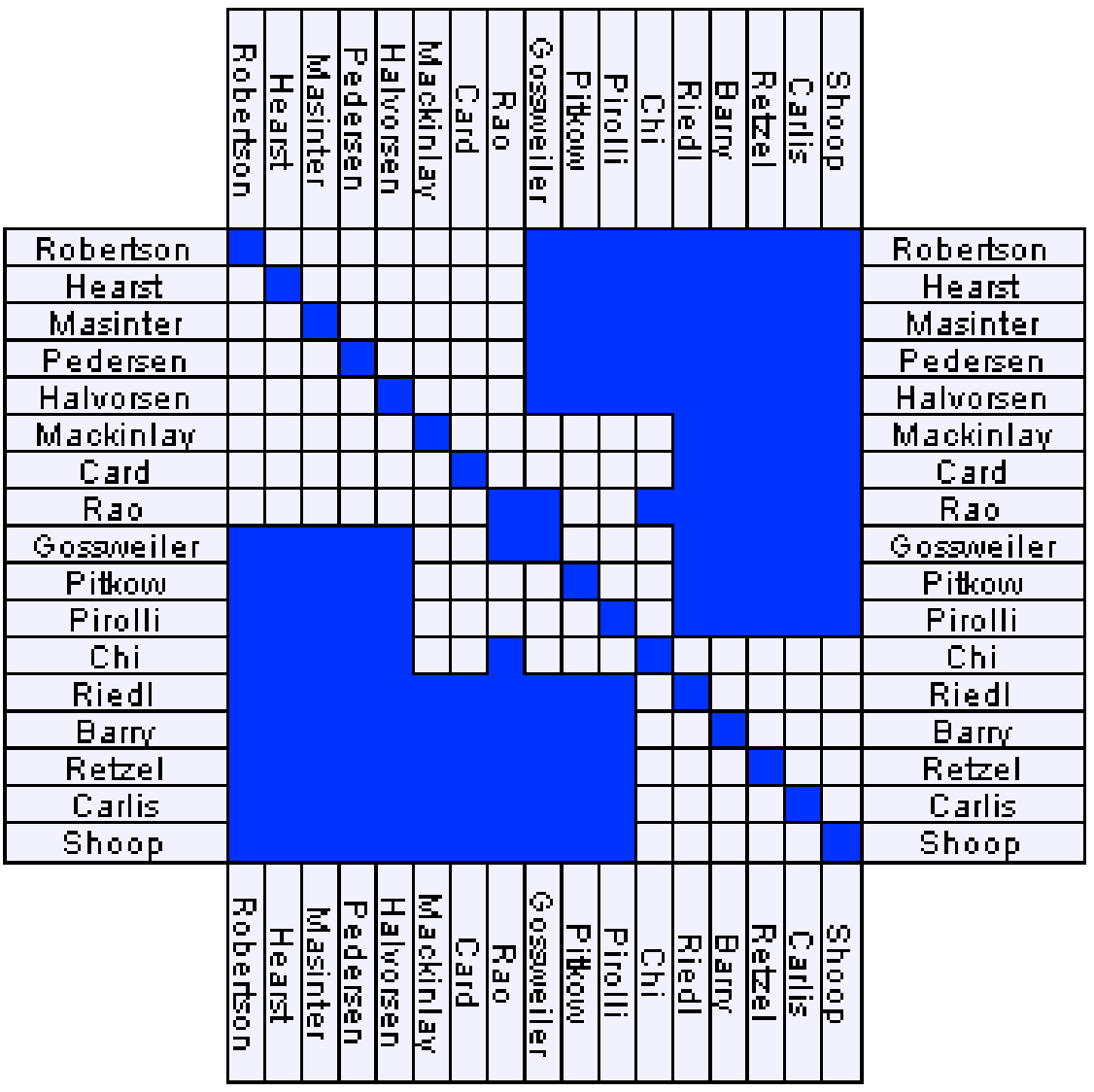}
\includegraphics[width=5cm]{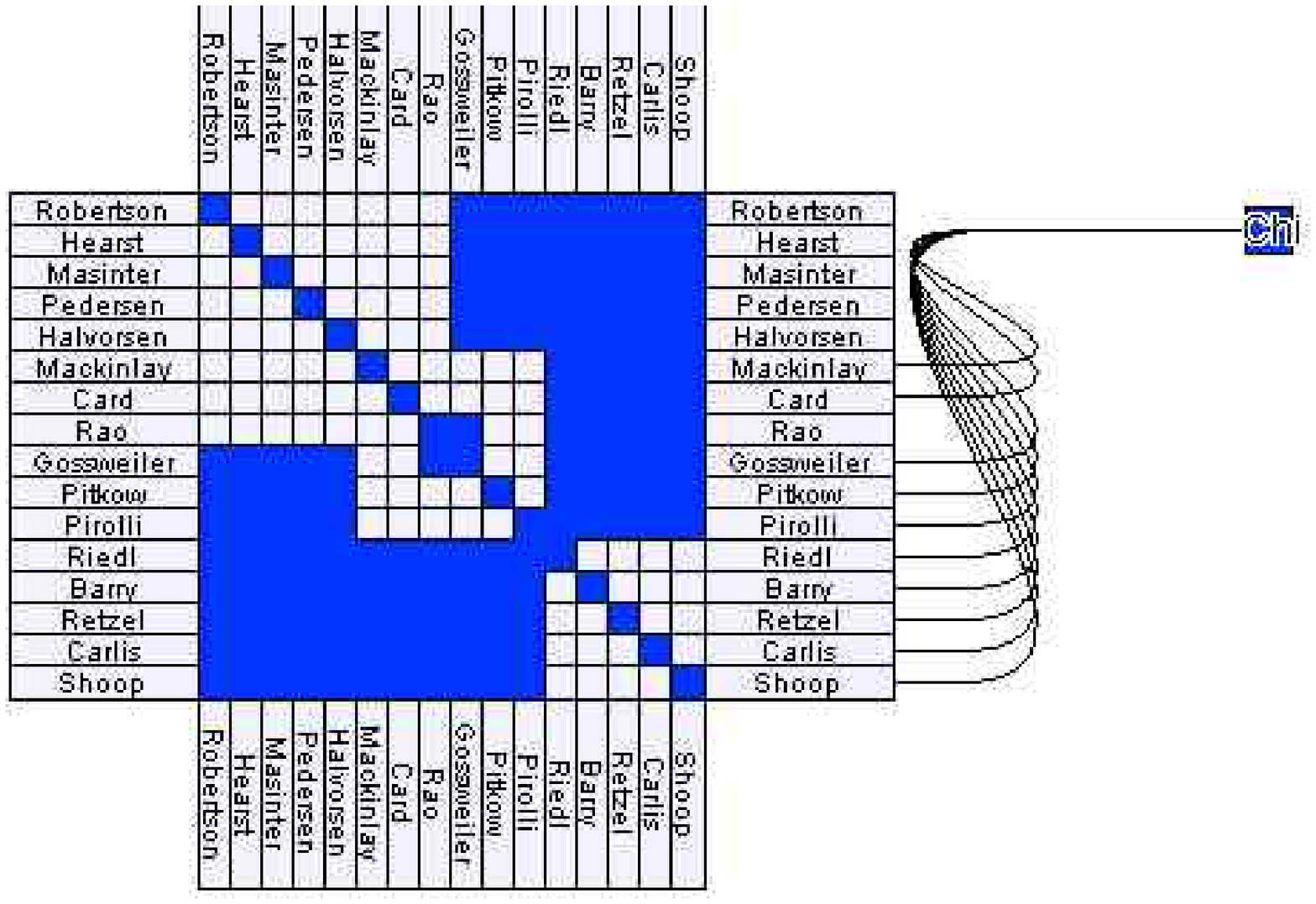}
\label{fig:inoutmatrix}	
\caption{Moving a node in and out of a matrix. The PARC community is shown on
the left, the influcence of Ed Chi outlined on the right figure.}
\end{figure}

While exploring the network, the interactions provided with NodeTrix
ease the analysis.  For example, moving an actor in and out of a
community (matrix) helps understand his influence on this community.
Figure~\ref{fig:inoutmatrix} illustrates this operation, showing that
if Ed Chi is extracted from the PARC community, then the community is
disconnected.  This operation also helps understand the matrix
representation for novice users, as they can drag out of the matrix
each actor, one at a time, visualizing his relations to the others and
comparing it with its matrix representation.

\subsection{Patterns of Collaboration}

The main result of our case study is the identification of different
collaboration patterns: cross patterns and block patterns.

Figure~\ref{fig:cross} reveals the collaboration pattern of Ben
Shneiderman, main actor of the InfoVis field.  This aggregated matrix
is very sparse and shows only a complete first row and first column.
We named this pattern a cross pattern because if Ben is placed
somewhere else in the matrix (as it is often the case when the matrix
is not reordered), the visible pattern is a large cross.  This pattern
reveals that Ben collaborates with all researchers in this matrix.
However, the low density shows that Ben's collaborators generally do
not work together: they are probably students he is supervising.
Figure~\ref{fig:infoviscompact} reveals several matrices with this
pattern of collaboration: Plaisant et al., Bederson et al. and, Eick
et al.

Figure~\ref{fig:clique} reveals the collaboration pattern of
researchers from Berkeley.  The aggregated matrix is almost a clique,
it is a very dense community.  Contrary to the previous pattern, this
one reveals that researchers strongly collaborate with each other and
not only a single one.  Figure~\ref{fig:infoviscompact} shows that
Parc has the same collaboration pattern.

Note that the community formed by Stephen Roth is to be placed in an
intermediate category (Figure~\ref{fig:intermediate}).  Stephen is
central in this community, but several blocks are visible, meaning
that researchers also collaborate with each other.

\begin{figure}[ht]
\centering
\includegraphics[width=2.5cm]{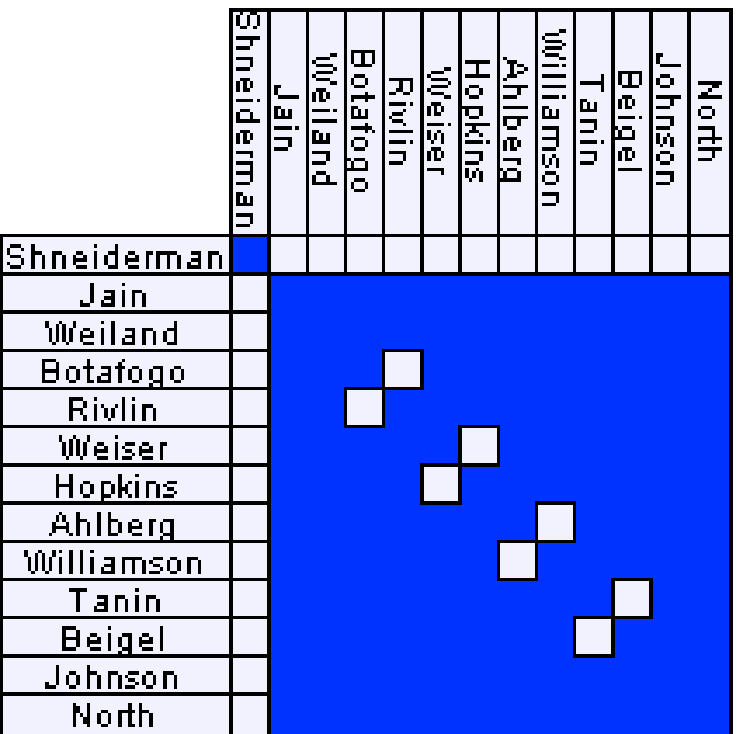}
\includegraphics[width=2.5cm]{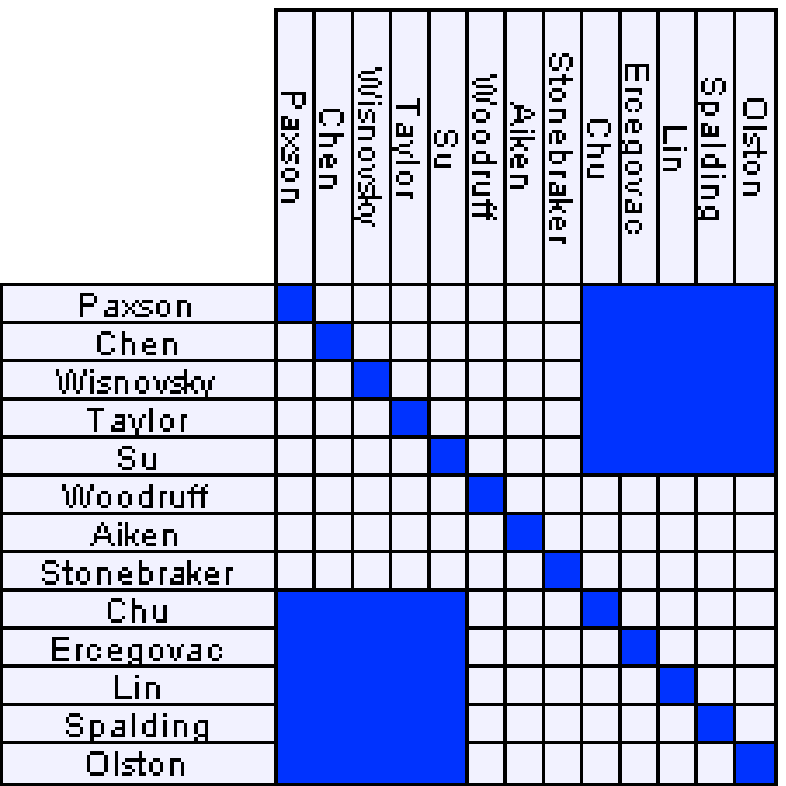}
\includegraphics[width=2.5cm]{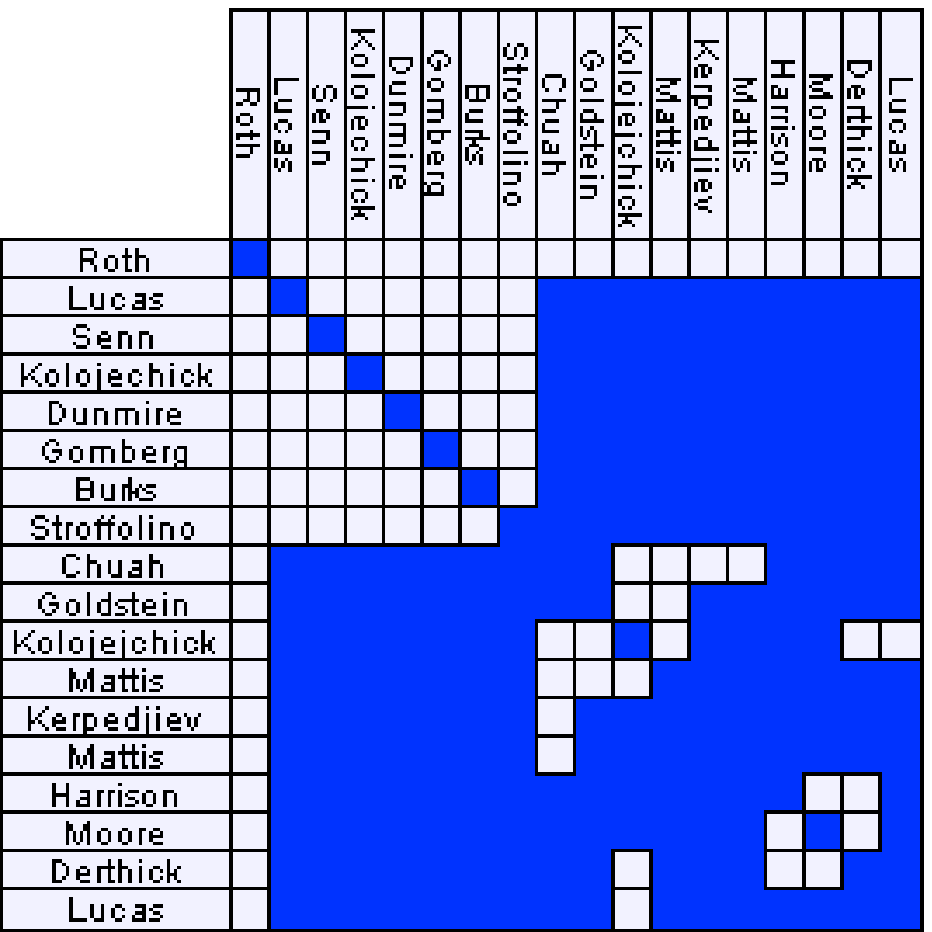}
\label{fig:collaborationpatterns}	
\caption{Three collaboration patterns (from left to right): (a)Cross-Pattern: 
Shneiderman and his collaborators, (b) Block-Pattern: Researchers at Berkeley,
(c) Mixed-Pattern: Roth and his collaborators}
\end{figure}

\section{Conclusion and Future Work}
We have introduced a novel visualization called \emph{NodeTrix}.  This
representation integrates the best of two traditional network
representations: node-link diagrams and adjacency matrix-based
representations.  The strength of this representation for analyzing
social networks is that it is combining the familiarity of node-link
diagrams to understand the global structure of the network and the
readability of matrices for detailed community analysis.

We described a set of interactions support the manipulation of
NodeTrix and help analyzing networks.  We also proposed an animation
to smooth the transition between node-link diagrams and matrices.  In
a slower mode, this animation can be used to help novice users
understand how matrices work.  Finally, we have illustrated the
effectiveness of NodeTrix with a case study of the InfoVis
publications data.  

We plan to extend our system in several directions and to perform
evaluations on its use with real analysts.  The interactive
capabilities of NodeTrix are well suited to collaborative analyzes so
an obvious extension include collaborative edition, either through the
network or in a shared environment with large displays.

We have iterated on several alternative representations to visualize
social networks and believe that NodeTrix is among the most effective
and simplest to understand.  We plan to release it soon as a component
of the InfoVis Toolkit (ivtk.sourceforge.net).


\bibliographystyle{abbrv}
\bibliography{nodetrixRR}

\end{document}